\documentclass[12pt,letterpaper]{article}
\usepackage{t1enc}
\usepackage[latin1]{inputenc}
\usepackage[english]{babel}
\usepackage{graphicx}
\usepackage{amsmath}
\usepackage{amssymb}
\renewcommand{\baselinestretch}{1.75}

\begin{document}

\title{Multimodal Distributions for Circular Axial Data}
\author{Fern\'andez-Dur\'an, J.J. and Gregorio-Dom\'inguez, M.M. \\
ITAM\\
E-mail: mercedes@itam.mx}
\renewcommand{\baselinestretch}{1.00}
\date{}
\maketitle

\renewcommand{\baselinestretch}{1.75}

\begin{abstract}
The family of circular distributions based on non-negative trigonometric sums (NNTS), developed by Fern\'andez-Dur\'an (2004), is highly flexible for modeling datasets exhibiting multimodality and/or skewness. In this article, we extend the NNTS family to axial data by identifying conditions under which the original NNTS family is suitable for modeling undirected vectors. Since the estimation is performed using maximum likelihood, likelihood ratio tests are developed for characteristics of the density function such as uniformity and symmetry, as well as to compare different axial populations through homogeneity tests. The proposed methodology is applied to real datasets involving orientations of rocks, animals, and plants.
\end{abstract}


\newpage


\section{Introduction}

Axial data, defined as data representing undirected vectors, are relevant in many applications across different disciplines, notably in Paleomagnetism for the statistical analysis of Earth's magnetic field directions recorded during rock formation (Onstott, 1980; Tauxe, 2010). Axial data also appear in analyses of astronomical body orientations (Hawley \& Peebles, 1975), protein structure axes (Barlow \& Thornton, 1988), and certain anatomical structures in animals and humans (Vrtovec et al., 2009), such as the posterior corneal curvature of the eye (Koch et al., 2012). In Ecology, axial data assist in tracking animal magnetic orientation (Batschelet, 1981; Wiltschko \& Wiltschko, 1995 and 2005; Fitak \& S\"onke, 2017) and in Environmental Science, they aid the study of vegetation orientation patterns, their relation to climate change (Sherratt, 2015), and associated effects on wind direction (Mingione et al., 2015).

Arnold \& Sengupta (2011) describe various methods to generate densities for an axial random variable, \(\theta \in [0, \pi)\), from either a linear random variable \(X\) or a circular random variable \(\phi \in [0,2\pi)\). The first method involves wrapping a linear random variable: \(\theta = X \mod \pi\). The second, called adding, defines the axial density as the sum of a circular density at \(\phi\) and \(\phi + \pi\), equivalent to wrapping a circular variable: \(\theta = \phi \mod \pi\). The third method uses polar transformation of a bivariate linear vector \((X,Y)\), defining \(\theta = \tan^{-1}(Y/X) + \pi/2\). Early studies on the analysis of axial data initially approached the problem by transforming the axial variable $\theta$ into a circular random variable $\phi$ through a simple doubling transformation, $\phi = 2\theta$ (Batschelet, 1981). The transformed data were then analyzed using standard models developed for circular data. This approach, commonly referred to as the method of angle doubling, produces models for axial data that are directly derived from corresponding models for circular data.

Common models for axial data include the axial von Mises (AvM), axial wrapped Cauchy (AWC), and sine-skewed distributions. The AvM model (Arnold \& SenGupta, 2006), derived using the adding method, has the density:
\[
f_{AvM}(\theta \mid \mu, \kappa) = \frac{1}{\pi I_0(\kappa)} \cosh(\kappa \cos(\theta - \mu)), \quad 0 \le \theta < \pi,
\]
where \(\mu\) is the location parameter, \(\kappa\) is the dispersion parameter, and \(I_0(\kappa)\) denotes the modified Bessel function of the first kind (order zero). Similarly, the AWC model, obtained by wrapping a linear Cauchy random variable, is symmetric and unimodal:
\[
f_{AWC}(\theta) = \frac{1 - \rho^4}{\pi(1 + \rho^4 - 2\rho^2 \cos(2(\theta - \mu)))}, \quad -\frac{\pi}{2} \le \theta < \frac{\pi}{2},
\]
where \(0 \le \rho \le 1\) and $\mu$ is the location parameter. Additionally, Arnold \& SenGupta (2011) introduced the angular central Gaussian (AACG) model, which generalizes the AWC by using a polar transformation of a bivariate Gaussian vector:
\[
f_{AACG}(\theta) \propto \left(1 + \alpha \sin\left(2\left(\theta - \mu + \frac{\pi}{4} - \beta\right)\right)\right)^{-1}, \quad 0 \le \theta < \pi,
\]
with parameters \(\alpha\), \(\beta\) and $\mu$. Abe et al. (2012) proposed sine-skewed (SineSk) axial models, introducing asymmetry through a perturbation term:
\[
f_{SineSk}(\theta) = (1 + \lambda \sin(k(\theta-\mu)))f_0(\theta-\mu), \quad -\frac{\pi}{2} \le \theta < \frac{\pi}{2},
\]
where \(f_0(\theta)\) is the base symmetric axial density, \(k=2\), and \(-1 \le \lambda \le 1\). When \(\lambda=0\), the SineSk model reduces to the symmetric base model. Multimodal axial densities may also be constructed as mixtures (e.g., mixtures of axial von Mises distributions), though this approach complicates numerical parameter estimation when the number of modes increases.

In this paper, we generalize the family of flexible circular NNTS distributions to axial data, enabling modeling of multimodal and asymmetric distributions beyond previous literature models. The maximum likelihood estimation (MLE) of NNTS axial model parameters is efficiently carried out using a numerical Newton optimization algorithm on manifolds. This facilitates direct application of likelihood ratio tests for assessing features such as uniformity, symmetry, or homogeneity across axial populations.

This paper is organized as follows. Section two presents conditions necessary to extend NNTS distributions for modeling axial data and outlines constraints required for symmetry. It also introduces likelihood ratio tests based on parameter restrictions. Section three demonstrates the proposed methodology with real datasets involving orientations from geology, animal behavior, and plant biology. Finally, section four provides concluding remarks.

\section{A Family of Distributions for Circular Axial Data Based on Nonnegative Trigonometric Sums}

\subsection{Definition}

The density function of a circular random variable \(\phi \in (0,2\pi]\), based on non-negative trigonometric sums (NNTS) with a complex parameter vector \(\underline{c}\), is defined as
\begin{equation}
f_{NNTS}(\phi \mid \underline{c})= \left|\left|\sum_{k=0}^{M}c_k e^{ik\phi} \right|\right|^2 = \sum_{k=0}^{M}\sum_{m=0}^{M} c_k\bar{c}_m e^{i(k-m)\phi},
\label{NNTSdensity}
\end{equation}
where \(i=\sqrt{-1}\), and \(c_k=c_{re,k}+ic_{im,k}\) are complex parameters for \(k=0,1,\dots,M\), with \(c_{re,k}\) and \(c_{im,k}\) being the real and imaginary parts, respectively. Here, \(e^{ik\phi}=\cos(k\phi) + i\sin(k\phi)\). The NNTS density is thus the squared norm of a complex sum. To define a valid density function, the parameters must satisfy
\begin{equation}
\sum_{k=0}^{M} ||c_k||^2 = \frac{1}{2\pi},
\label{NNTSdensityconstraint}
\end{equation}
where \(||c_k||^2 = c_{re,k}^2 + c_{im,k}^2\). To ensure identifiability of the $c$ parameters, we constrain $c_0$ to be a positive real number, which immediately implies that $c_{im,0} = 0$. From Equation \ref{NNTSdensityconstraint}, it then follows that $c_0 = \frac{1}{2\pi} - \sum_{k=0}^{M} \|c_k\|^2$.

A density function for an axial random variable \(\theta \in [0,\pi)\) must satisfy the condition \(f(\theta)=f(\theta \pm \pi)\) since an axis represents an undirected unit vector. If \(\phi\) is the angle of a unit vector on a circle, then the density should have identical values at \(\phi\) and \(\phi \pm \pi\), representing opposite directions. Thus, the condition for the NNTS circular density becomes
\begin{equation}
f_{NNTS}(\phi \mid \underline{c}) = f_{NNTS}(\phi \pm \pi \mid \underline{c}).
\label{NNTSaxialcondition}
\end{equation} 

Evaluating this condition:
\begin{eqnarray*}
f_{NNTS}(\phi \pm \pi \mid \underline{c}) & = & \sum_{k=0}^{M}\sum_{m=0}^{M} c_k\bar{c}_m e^{i(k-m)(\phi \pm \pi)} \\
& = & \sum_{k=0}^{M}\sum_{m=0}^{M} (c_ke^{i(\pm k \pi)})(\bar{c}_me^{-i(\pm m \pi)}) e^{i(k-m)\phi} \\
& = & \sum_{k=0}^{M}\sum_{m=0}^{M} c_k^* \bar{c}_m^* e^{i(k-m)\phi}.
\end{eqnarray*}
Since \(e^{i(\pm k \pi)}=\cos(k \pi)\), and \(\cos(k \pi)=1\) for even \(k\) and \(\cos(k \pi)=-1\) for odd \(k\), the condition \(f_{NNTS}(\phi)=f_{NNTS}(\phi\pm\pi)\) implies:
\[c_k^*=c_k \text{ for even } k \quad\text{and}\quad c_k^*=-c_k \text{ for odd } k.\]
To satisfy these conditions, we must have \(c_k=0\) for all odd \(k\). Thus, the NNTS density for an axial random variable \(\theta \in [0,\pi)\), with a complex parameter vector \(\underline{v}=(v_0,v_1,\dots,v_M)^T\), is defined by considering only even indexes in the summation of Equation \eqref{NNTSdensity}:
\begin{equation}
f_{NNTSaxial}(\theta \mid \underline{v}) = \left|\left|\sum_{k=0}^{M}v_k e^{i2k\theta}\right|\right|^2 = \sum_{k=0}^{M}\sum_{m=0}^{M} v_k\bar{v}_m e^{i2(k-m)\theta}.
\label{NNTSaxialdensity}
\end{equation}

This corresponds to doubling the original circular angle \(\phi\), a common technique for analyzing axial data. For this to define a valid density, the parameters must satisfy
\begin{equation}
\sum_{k=0}^{M} ||v_k||^2 = \frac{1}{\pi},
\label{NNTSdensityaxialconstraint}
\end{equation}
and, to ensure identifiability of the $v$ parameters, we constrain $v_0$ to be a positive real number. From Equation \ref{NNTSdensityaxialconstraint}, it then follows that
$v_0 = \frac{1}{\pi} - \sum_{k=1}^{M} \|v_k\|^2$. Alternatively, the NNTS axial density can be expressed as:
\begin{equation}
f_{NNTSaxial}(\theta \mid \underline{v}) = \frac{1}{\pi}\sum_{k=0}^{M}\sum_{m=0}^{M} v_k\bar{v}_m e^{i2(k-m)\theta},
\label{NNTSaxialdensityuniform}
\end{equation}
with \(\sum_{k=0}^{M} ||v_k||^2 = 1\), placing parameters on the unit complex hypersphere \(CS^{M}=\{\underline{v} : \sum_{k=0}^{M} ||v_k||^2 = 1\}\) and $v_0$ being a positive real number.

The number of free parameters in \(\underline{v}\) equals \(2M\), and \(M\) determines the maximum number of modes for the NNTS axial density. The support of the axial random variable can be \((0,\pi]\) due to $f_{NNTSaxial}(\theta \mid \underline{v}) = f_{NNTSaxial}(\theta \pm \pi \mid \underline{v})$, although the interval \([-\frac{\pi}{2},\frac{\pi}{2})\) is also common in literature. Note that NNTS axial models are nested; models with smaller \(M\) are special cases of those with larger \(M\), facilitating likelihood ratio tests. Specifically, the case \(M=0\) corresponds to the uniform axial density.

The trigonometric moments, $E[e^{ir\theta}]$, for an odd $r$ are equal to zero. This is because, for a circular axial random variable,
\[
E[e^{ir\theta}] = E[e^{ir(\theta+\pi)}] = e^{ir\pi} E[e^{ir\theta}] = \cos(r\pi) E[e^{ir\theta}] = (-1)^r E[e^{ir\theta}],
\]
which equals $-E[e^{ir\theta}]$ since $r$ is odd. Therefore, $E[e^{ir\theta}] = 0$ for odd $r$. For even $r$, the trigonometric moments of the NNTS axial distribution in Equation \eqref{NNTSaxialdensityuniform} are given by

\begin{equation}
E(e^{ir\theta}) = \frac{1}{\pi} \int_{0}^{\pi} e^{ir\theta} \sum_{k=0}^{M} \sum_{m=0}^{M} v_k \bar{v}_m e^{i2(k-m)\theta} \, d\theta 
= \frac{1}{\pi} \sum_{k=0}^{M} \sum_{m=0}^{M} v_k \bar{v}_m \int_{0}^{\pi} e^{i(2(k-m)+r)\theta} \, d\theta.
\end{equation}
Since \(k\) and \(m\) are integers and \(r\) is even, \(2(k - m) + r\) is also even. Therefore, the integral \(\int_{0}^{\pi} e^{i(2(k - m) + r)\theta} \, d\theta\)
equals zero when \(2(k - m) + r \neq 0\), and equals \(\pi\) when \(2(k - m) + r = 0\).
Thus, for even \(r\), the trigonometric moment simplifies to:
\begin{equation}
E(e^{ir\theta}) = \sum_{m=\frac{r}{2}}^{M} v_{m-\frac{r}{2}} \bar{v}_m
\end{equation}
for \(r = 2, 4, \ldots, M\) when \(M\) is even, and \(r = 2, 4, \ldots, M - 1\) when \(M\) is odd. From these trigonometric moments, one can derive other characteristics of the NNTS axial distribution, such as the mean direction, circular variance, asymmetry coefficient, and kurtosis (see Mardia \& Jupp, 2000). Moreover, the invariance property of the maximum likelihood estimation method can be used to obtain estimates of functions of the parameters, such as the trigonometric moments and other derived quantities.

In many practical applications, it is necessary to model axial data that are symmetric with respect to an angle \(\mu \in (0, \pi]\), known as the axis of symmetry. A NNTS axial model is symmetric if the complex parameter vector \(\underline{v}\) is restricted to be real and nonnegative (see Fern\'andez-Dur\'an \& Gregorio-Dom\'inguez, 2025). The corresponding density function is defined as:
\begin{equation}
f_{NNTSaxialsym}(\theta \mid \underline{v}_R) = \frac{1}{\pi} \left|\left| \sum_{k=0}^{M} v_{Rk} e^{i2k(\theta - \mu)} \right|\right|^2 
= \frac{1}{\pi} \sum_{k=0}^{M} \sum_{m=0}^{M} v_{Rk} v_{Rm} e^{i2(k-m)(\theta - \mu)}.
\end{equation}
Here, the vector \(\mathbf{v}\) is restricted to be a nonnegative real vector, with the parameter space defined as \(\{\mathbf{v} : v_{R0} > 0,\; v_{Rk} \ge 0 \text{ for } k = 1, \ldots, M,\; \sum_{k=0}^{M} \|v_{Rk}\|^2 = 1\}\). Alternatively, the complex vector \(\underline{v}^*\) can be defined with \(v_k^* = v_{Rk} e^{-i2k\mu}\). The total number of free parameters in a symmetric NNTS axial model is \(M + 1\).

Since the proposed NNTS axial density can be expressed as a quadratic form, 
\[
\frac{1}{\pi} \underline{v}^H \underline{e} \underline{e}^H \underline{v},
\]
with \(\underline{v} = (v_0, v_1, \ldots, v_M)^{T}\) denoting the parameter vector, \(\underline{v}^{H}\) its Hermitian (conjugate transpose), \(\underline{e} = (1, e^{i2\theta}, e^{i4\theta}, \ldots, e^{i2M\theta})^{T}\) the vector of trigonometric statistics, and \(\underline{e}^{H}\) its Hermitian, the estimation algorithm described by Fern\'andez-Dur\'an \& Gregorio-Dom\'inguez (2010) can be applied. This algorithm involves a modified Newton method on the complex unit hypersphere parameter space (manifold), 
which can be locally approximated by a hyperplane known as the tangent space. 
The modified Newton algorithm of Fern\'andez-Dur\'an \& Gregorio-Dom\'inguez (2010) operates by searching for maxima in directions defined within the tangent space, 
using the complex gradient vector and the Hessian matrix of the log-likelihood function, and then reprojecting (retracting) onto the complex unit hypersphere. 

For the symmetric NNTS axial model, at each iteration of the modified Newton algorithm, the norms of the elements of the complex vector \(\mathbf{v}^*\) are computed, 
and the axis of symmetry, \(\mu\), is optimized over a grid of values in the interval \([0, \pi)\). 
After convergence of the modified Newton algorithm, the estimates of the nonnegative real parameter vector \(\mathbf{v}\) and the axis of symmetry \(\mu\) are obtained.

Since the model parameters are estimated via maximum likelihood, likelihood ratio tests can be defined as
\[
LLR_{RG} = -2 \left( \hat{\ell}(\hat{\underline{v}}_R \mid \underline{\theta}) - \hat{\ell}(\hat{\underline{v}}_G \mid \underline{\theta}) \right),
\]
where \(\underline{\theta}\) is the vector of observed axial data, \(\hat{\underline{v}}_G\) is the MLE under the general model, \(\hat{\underline{v}}_R\) is the MLE under the restricted model, and \(\hat{\ell}(\cdot)\) denotes the corresponding maximized log-likelihood values. Under standard regularity conditions, \(LLR_{RG}\) asymptotically follows a chi-squared distribution with degrees of freedom equal to the number of constraints imposed to derive \(\underline{v}_R\) from \(\underline{v}_G\).

For instance, a test for symmetry can be performed by comparing the general NNTS axial model to a symmetric version. In this case, the test statistic follows a chi-squared distribution with \(M - 1\) degrees of freedom. To test for homogeneity across \(P\) different populations with independent samples, one compares the sum of the maximized log-likelihoods from each population to the log-likelihood obtained from pooling all data into a single model. The degrees of freedom for the test are given by:
\[
\sum_{k=1}^{P} 2M_k - 2M_{pooled},
\]
where \(2M_k\) is the number of parameters in the fitted model for population \(k\), and \(2M_{pooled}\) is the number of parameters in the pooled model.
Under the null hypothesis of homogeneity of the $P$ populations, and given the nested structure in the parameter $M$ of the NNTS axial models, we have $M_k = M_{\text{pooled}}$ for $k = 1, 2, \ldots, P$. Consequently, the number of degrees of freedom of the chi-squared likelihood ratio test for homogeneity simplifies to $2(P-1)M_{\text{pooled}}$. 
In practical applications of the homogeneity test, the researcher may know the value of $M_{\text{pooled}}$ from theoretical considerations or previous studies. 
If this is not the case, the researcher can apply the homogeneity test for different values of $M_{\text{pooled}}$ and calculate a corrected final $p$-value from the $p$-values of the tests corresponding to these different values. Various methods for correcting $p$-values due to multiple hypothesis testing can be used (Goeman \& Solari, 2014; Cinar \& Viechtbauer, 2022).
In particular, we considered the Bonferroni and Tippett corrections for multiple testing, with the adjusted \( p \)-values defined as
\[
p_{c}^{(B)} = \min\left\{1, \min(p_1, p_2, \ldots, p_P) \times P\right\}
\quad \text{and} \quad
p_{c}^{(T)} = 1 - (1 - \min(p_1, p_2, \ldots, p_P))^{P},
\]
respectively, where \( p_{c}^{(B)} \) and \( p_{c}^{(T)} \) denote the Bonferroni- and Tippett-corrected \( p \)-values, and \( p_1, p_2, \ldots, p_P \) are the individual \( p \)-values from the \( P \) tests.

\section{Real Data Examples}

\subsection{Feldspar Lath Orientations in Basalt}

\begin{figure}[h]
\center{
\includegraphics[scale=0.7, bb= 0 0 504 504]{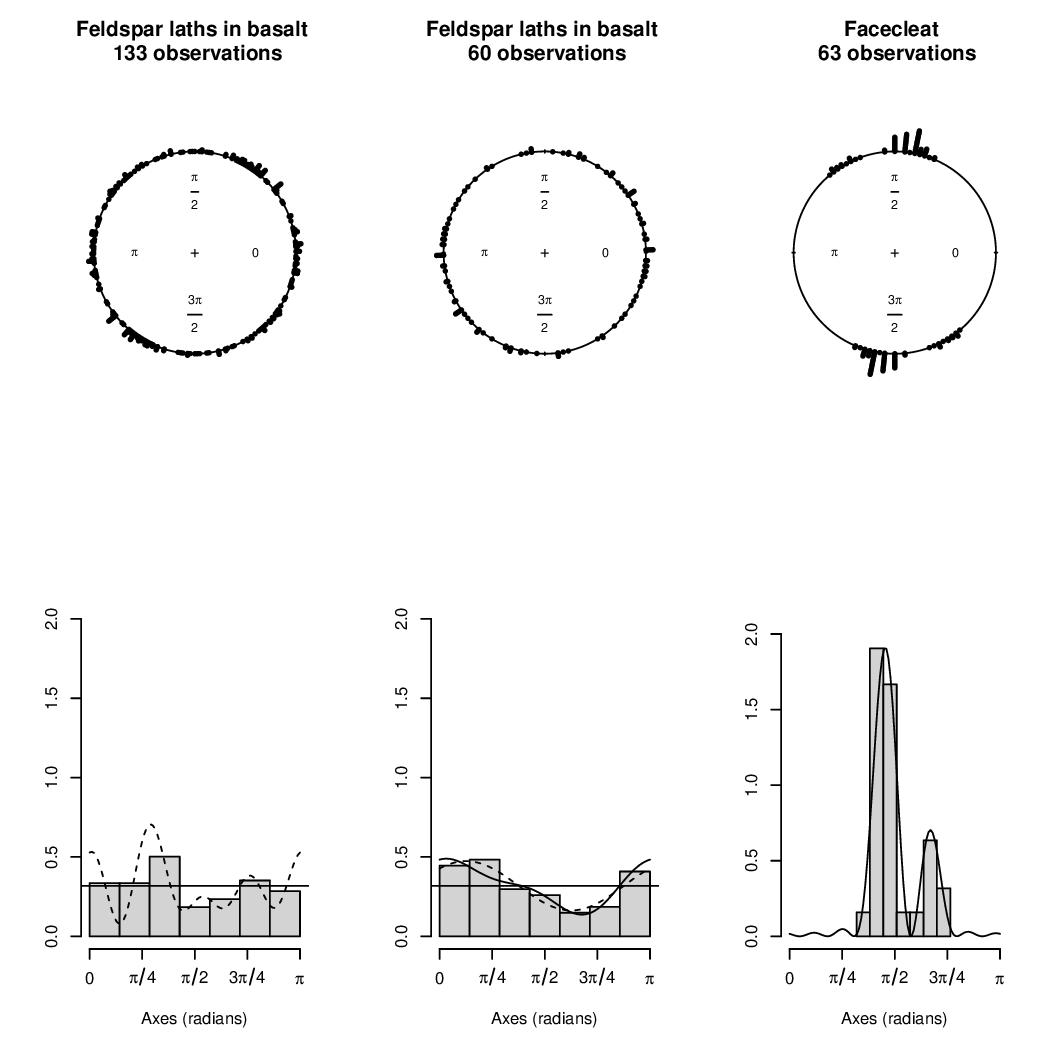}
\caption{Feldspar laths in basalt and face cleat in a coal seam datasets: Circular dot plot and histogram with best BIC-fitted NNTS axial model (solid line) and best AIC-fitted NNTS axial model (dashed line) for the feldspar laths in basalt dataset with 133 observations (left plot), the feldspar laths in basalt dataset with 60 observations (center plot), and the face cleat dataset with 63 observations (right plot). For the center plot, a second solid curve is included, corresponding to the fitted NNTS axial asymmetric model with $M$=2.
\label{feldspar0graphthreetogether}
}
}
\end{figure}

Feldspar crystals within rocks often align in parallel or subparallel patterns, providing valuable insights into the processes of rock formation and historical geological events, such as volcanic eruptions. Dataset B2 from Fisher (1993) contains measurements of the long-axis orientations of 133 feldspar laths in basalt. Table \ref{feldspar0tablethreetogether} (columns 2 to 4) presents maximized log-likelihood, Bayesian Information Criterion (BIC), and Akaike Information Criterion (AIC) values for fitted NNTS axial models with parameter $M$ ranging from 0 to 13. Figure \ref{feldspar0graphthreetogether} (left plot) shows a circular dot plot and histogram of the dataset along with the best-fitted NNTS axial densities selected by BIC (\(M=0\)) and AIC (\(M=4\)). A uniformity test by Fern\'andez-Dur\'an \& Gregorio-Dom\'inguez (2024), applied by doubling the axial angle, rejects the null hypothesis of uniformity implied by the BIC-selected model (\(M=0\)) in favor of the multimodal NNTS axial model chosen by AIC (\(M=4\)), yielding a $p$-value of less than 0.01.

Dataset B5 from Fisher (1993) contains orientations of an additional 60 feldspar laths in basalt. Fisher emphasized the importance of testing uniformity and determining the number of modes in this dataset. Table \ref{feldspar0tablethreetogether} (columns 5 to 7) presents maximized log-likelihood, Akaike Information Criterion (AIC), and Bayesian Information Criterion (BIC) values for NNTS axial models with parameter \(M\) ranging from 0 to 6. Figure \ref{feldspar0graphthreetogether} (center plot) shows the histogram along with the best-fitting model according to BIC, which corresponds to a uniform axial density ($M=$0), and the best-fitting model according to AIC, which corresponds to a unimodal and symmetric density ($M=$1).

We conducted a likelihood ratio test comparing a symmetric NNTS model with \(M=1\) against an asymmetric NNTS model with \(M=2\), depicted as a dashed line in Figure \ref{feldspar0graphthreetogether} (center plot). The resulting $p$-value was 0.6077, indicating insufficient evidence to reject the simpler symmetric model (\(M=1\)) in favor of the asymmetric alternative (\(M=2\)).

Additionally, the uniformity test by Fern\'andez-Dur\'an \& Gregorio-Dom\'inguez (2024) produced a $p$-value between 0.01 and 0.05, leading to the rejection of the null hypothesis of uniformity at a 5\% significance level when comparing the uniform density model (\(M=0\)) with the best AIC-selected NNTS axial model with \(M=1\). Our results are consistent with those of Fisher (1993), supporting both the non-uniformity and unimodality of the dataset.

Table~\ref{feldsparpooled} presents the results of applying the NNTS axial homogeneity tests to the two populations of feldspar laths in basalt (with 133 from dataset~B2 and 60 from dataset~B5, respectively, both from Fisher,~1993) for values of \( M_{\text{pooled}} \) ranging from 1 to 6. The table reports the maximized log-likelihood, BIC, and AIC values of the models fitted to the pooled population, as well as the \( p \)-values from the NNTS axial homogeneity tests. For \( M_{\text{pooled}} = 1, \ldots, 6 \), the null hypothesis of homogeneity is not rejected at the 5\% significance level. The combined \( p \)-values, obtained by aggregating the six homogeneity tests over \( M_{\text{pooled}} = 1, \ldots, 6 \), are equal to 1.0000 under the Bonferroni correction and 0.6752 under the Tippett correction, indicating no evidence against the null hypothesis of homogeneity.

\subsection{Face Cleat in a Coal Seam}

Face cleats are longitudinal fractures that represent significant geological features within coal seams-structures composed of coal deposits embedded within layers of rock. Variations in the orientation of these fractures can indicate hazardous mining conditions, including the potential presence of gas. Dataset B22 from Fisher (1993) includes 63 median directions of face cleats measured at the Wallsend Borehole Colliery in New South Wales, Australia. The median directions were measured at intervals of 20 meters along the coal seam. Consequently, the dataset exhibits sequential ordering, implying that the observations may not be independent. Despite the presence of potential dependence structures within the dataset, we analyzed the data as if it were a random sample to facilitate a direct comparison with results presented by Arnold \& SenGupta (2006).

Table \ref{feldspar0tablethreetogether} (columns 8 to 10) provides maximized log-likelihood, Akaike Information Criterion (AIC), and Bayesian Information Criterion (BIC) values for fitted NNTS axial models with parameter \(M\) ranging from 0 to 7. Figure \ref{feldspar0graphthreetogether} (right plot) shows the histogram along with the best-fitting AIC and BIC models, both of which select \(M=6\). The selected NNTS model with \(M=6\) displays two prominent modes within the observed range of data. This result contrasts with findings by Arnold \& SenGupta (2006), who considered only unimodal densities for the same dataset.

\subsection{Orientations of Termite Mounds}

\begin{figure}[h]
\center{
\includegraphics[scale=.9, bb= 0 0 504 504]{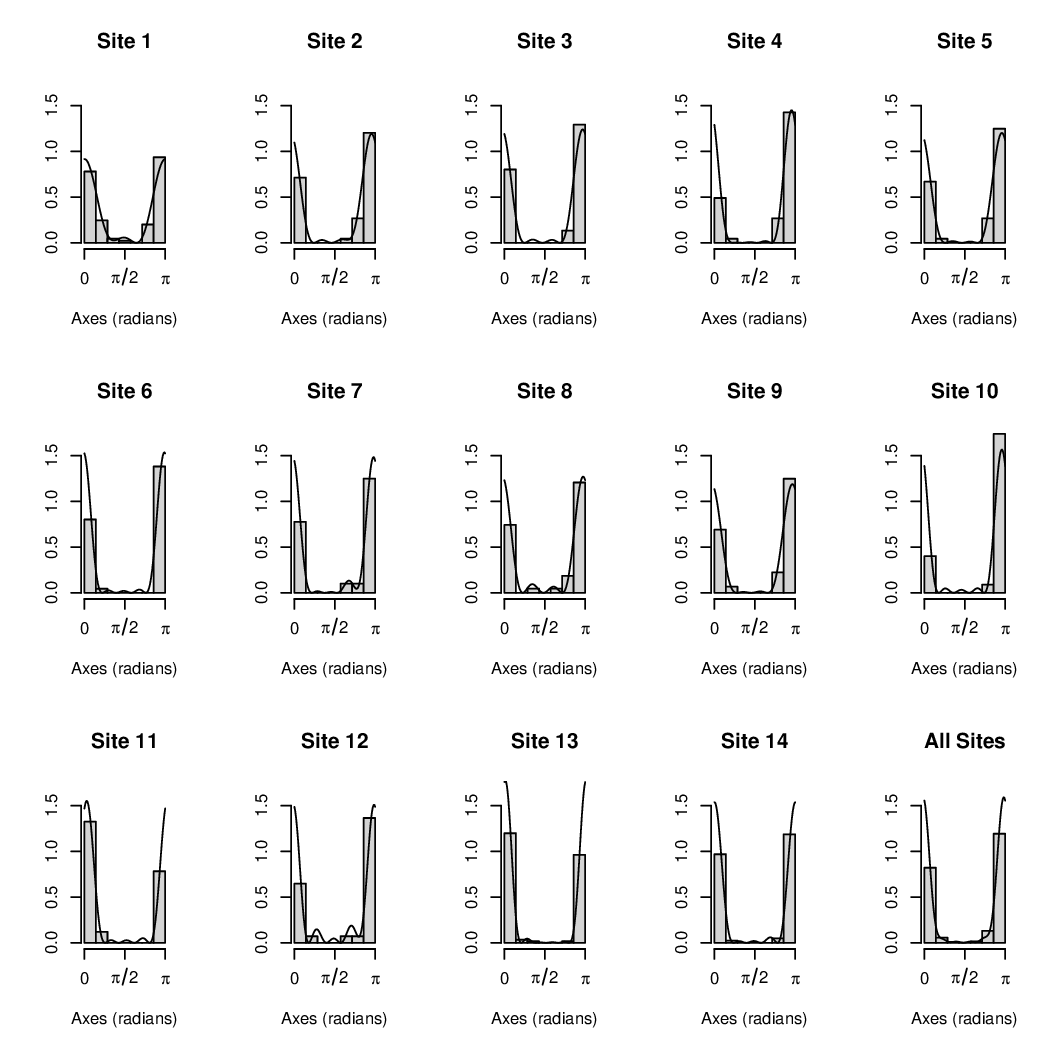}
\caption{Termite mounds dataset: Histograms and best BIC-fitted NNTS axial densities for each of the 14 individual sites. The final plot shows the histogram and best BIC-fitted NNTS axial density for the combined dataset from all sites.
\label{termitesgraph}
}
}
\end{figure}

Dataset B13 from Fisher (1993) contains measurements of termite mound orientations for the species \emph{Amitermes laurensis}, collected from 14 sites on the Cape York Peninsula, North Queensland, Australia. Observed orientations were standardized to the interval between 0 and \(\pi\) by subtracting \(\pi\) from any measurement greater than \(\pi\).
Figure \ref{termitesgraph} displays histograms for the orientation data at each of the 14 sites, along with sample sizes and geographical coordinates (latitude and longitude in degrees). The figure also shows the best-fitting NNTS axial densities according to the Bayesian Information Criterion (BIC) for each site. The final plot in Figure \ref{termitesgraph} presents the histogram and the best BIC-fitted NNTS axial density for the combined data from all 14 sites, which serves as the basis for a likelihood ratio test of homogeneity, with the null hypothesis stating that the NNTS axial density is the same across all 14 sites.
Table \ref{termitestable} provides maximized log-likelihood, AIC, and BIC values for NNTS axial densities with \(M = 0, 1, \ldots, 8\).

Table~\ref{termitespooled} (column~5) reports the \( p \)-values from the likelihood ratio tests of homogeneity based on NNTS axial models for the 14 individual sites, with \( M_{\text{pooled}} \) ranging from 1 to 8. For example, comparing the fitted NNTS models with \( M = 5 \) at each site against a joint model with \( M_{\text{pooled}} = 5 \) yields a \( p \)-value of 0.0062 with 130 degrees of freedom (see column 10 in Table \ref{termitestable}). For the cases where \( M_k = M_{\text{pooled}} = 4, 5, 6, \) and \( 7 \) for \( k = 1, 2, \ldots, 14 \), the null hypothesis of homogeneity across the 14 sites is rejected at the 5\% significance level, indicating that the orientation densities of termite mounds vary with geographic location (latitude and longitude). The multiple-testing-corrected \( p \)-values, considering the eight tests, are 0.0400 under the Bonferroni correction and 0.0393 under the Tippett correction, both leading to rejection of the null hypothesis of homogeneity at the 5\% significance level.

These findings confirm the results obtained by Fisher (1993),
who performed separate tests for the equality of mean directions and dispersions using a von Mises density model by doubling the axial angles to analyze them as circular data. However,
it should be noted that the von Mises model is inherently unimodal and therefore unsuitable for modeling multimodal datasets, such as several of those shown in Figure \ref{termitesgraph}, particularly evident in the histograms and fitted NNTS axial densities for site 8.

Fisher~(1993) conducted homogeneity tests specifically for the pairs of sites~5 and~14, and sites~6 and~8. Using a von Mises density model and doubling the axial angles to treat them as circular data, Fisher rejected the null hypothesis of equal mean directions for both pairs. However, for sites~6 and~8, Fisher (1993) concluded that the hypothesis of equal dispersions could not be rejected.
To enable a direct comparison with Fisher's~(1993) analysis, we performed NNTS axial likelihood ratio homogeneity tests for the same site pairs (5 and~14, and~6 and~8), using NNTS axial models with \( M_{\text{pooled}} \) ranging from~1 to~8. The resulting \( p \)-values, shown in Table~\ref{termitespooled} (columns~9 and~13), were all greater than~0.05, with a minimum of~0.4428 for sites~5 and~14 and a minimum of~0.7895 for sites~6 and~8. Consequently, the null hypothesis of homogeneity was not rejected for either pair, even when considering the Bonferroni- and Tippett-corrected \( p \)-values. These findings differ from those reported by Fisher~(1993), as Fisher's tests separately assessed equality of mean directions and dispersions, whereas our approach applies a unified likelihood ratio test for overall homogeneity.

\subsection{Magnetic Orientations of Ruminants}

Using satellite imagery, deer bedding patterns in snow, and field observations, Begall et al. (2008) analyzed the body orientations of domestic cattle, red deer, and roe deer during grazing and resting, with respect to true (geographic) North, at various locations worldwide with differing magnetic declinations. Their analysis suggested that the primary factor influencing the animals' orientations is alignment with Earth's magnetic field, even after accounting for factors such as site-specific declination, wind direction, solar position, and temperature. For each of the three samples, Begall et al. (2008) clearly rejected the null hypothesis of uniformity using the Rayleigh uniformity test. We utilized the dataset from Begall et al. (2008), consisting of mean axial vector directions for herds observed at 308 locations for domestic cattle, 40 locations for red deer, and 201 locations for roe deer. Using these data, we fitted NNTS axial models and conducted homogeneity tests on the orientations across the three populations. Particular attention was paid to assessing homogeneity between red and roe deer populations.

Table \ref{rumianttable} shows maximized log-likelihood, AIC, and BIC values for fitted NNTS axial models with different values of \(M\) (from 0 to 6 for red deer, and from 0 to 10 for cattle and roe deer). 

For testing homogeneity, Table~\ref{rumianttable2} presents the maximized log-likelihood, BIC, and AIC values for the NNTS models fitted to the pooled populations of cattle, red deer, and roe deer (columns~2--5), and to the pooled populations of red and roe deer (columns~6--9), for \( M_{\text{pooled}} \) ranging from~1 to~10. 
For the three-population case (cattle, red deer, and roe deer), all individual \( p \)-values (column~5) are well below~0.05, rejecting the null hypothesis of homogeneity, except for \( M_{\text{pooled}} = 1 \), which yields a \( p \)-value of~0.0731. In contrast, for the two-population case (red and roe deer), all individual \( p \)-values (column~9) are well above~0.05, indicating that the null hypothesis of homogeneity cannot be rejected at the 5\% significance level. 
For the three-population comparison, the Bonferroni- and Tippett-corrected \( p \)-values are \( 2.13\times10^{-19} \) and \( <0.0001 \), respectively. For the two-population comparison (red and roe deer), the corresponding corrected \( p \)-values are~1.0000 and~0.9999, respectively.

\begin{figure}[h]
\center{
\includegraphics[scale=0.7, bb= 0 0 504 504]{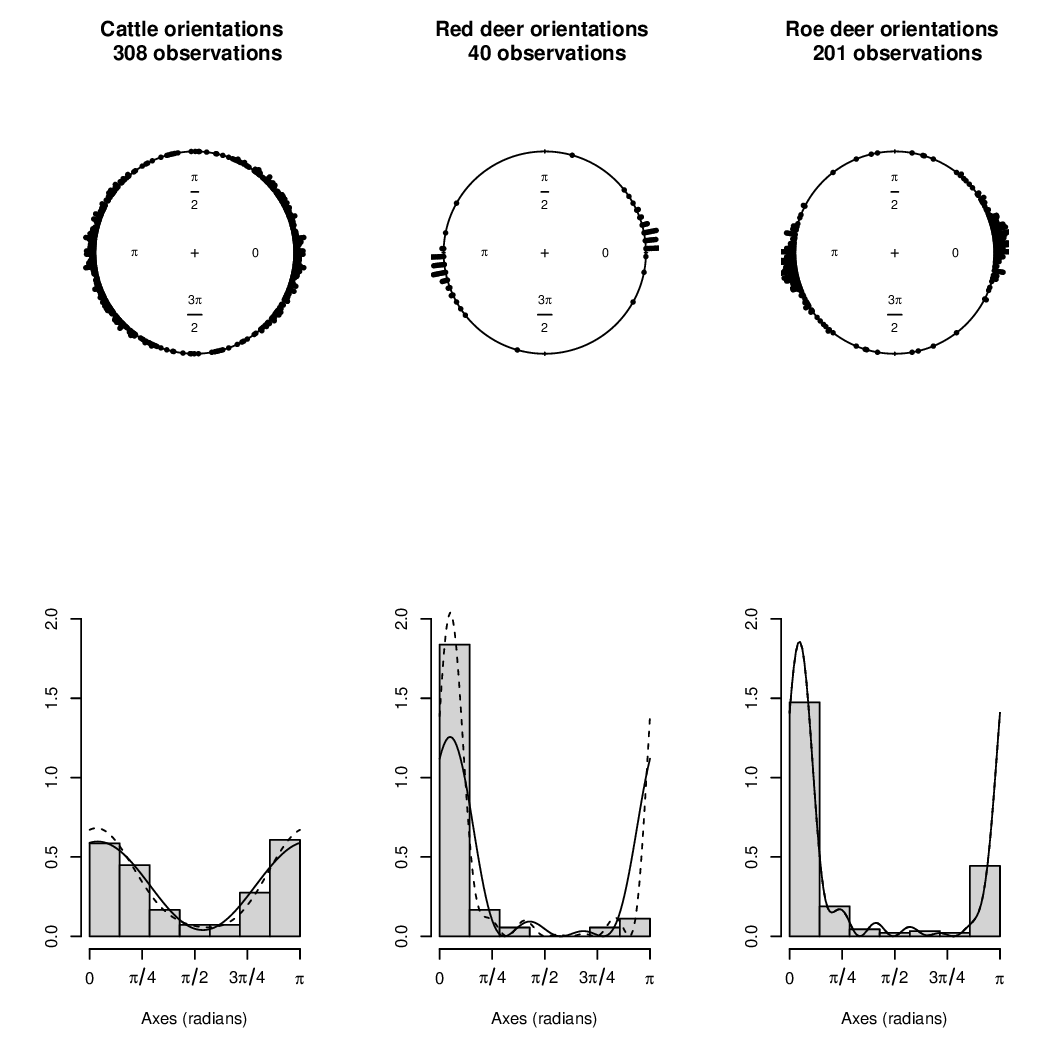}
\caption{Ruminant magnetic orientations: Circular dot plot and histogram with best BIC-fitted (solid line) and best AIC-fitted (dashed line) NNTS axial densities for the the domestic cattle dataset with 308 observations (left plot), the red deer dataset with 40 observations (center plot), and the roe deer dataset with 201 observations (right plot).
\label{begallanimalmagnetic}
}
}
\end{figure}

Figure \ref{begallanimalmagnetic} displays circular dot plots, histograms, and fitted density curves based on the best BIC and AIC NNTS axial models for cattle, red deer, and roe deer. 
For each population, the modes correspond to the average magnetic declination of the various sampling locations worldwide, as reported by Begall et al. (2008).

\subsection{Leaf Inclination Angles}

The spatial orientation of plant leaves, particularly leaf inclination angles, provides essential information regarding key biological processes such as photosynthesis efficiency, reflectance, temperature regulation, and overall resource utilization. Leaf inclination angle distributions vary according to plant species and geographic location. Recent advancements in digital imaging technology have significantly increased the availability of leaf inclination angle measurements, as algorithms can now efficiently determine these angles directly from digital images (see Pisek \& Adamson, 2020).

In botanical studies, the leaf inclination angle \(\theta_{leaf}\) typically ranges from 0 to \(\frac{\pi}{2}\). It is common practice in the literature to model the transformed linear variable \(X = \frac{2\theta_{leaf}}{\pi}\) using a Beta distribution with parameters \(\alpha\) and \(\beta\), and a probability density proportional to \(x^{\alpha -1}(1-x)^{\beta - 1}\). Based on parameter estimates for different plant species, de Wit (1965) proposed six theoretical distributions for leaf orientation angles: planophile (leaves predominantly horizontal), erectophile (leaves predominantly vertical), plagiophile (leaves predominantly inclined), extremophile (leaves predominantly horizontal and vertical), spherical (leaves oriented isotropically like a sphere), and uniform. The parameters of the Beta distribution for the plagiophile case satisfy $\alpha=\beta$ with $\alpha,\beta>1$, whereas for the extremophile case, they satisfy $\alpha=\beta$ with $\alpha,\beta<1$, indicating that the Beta distribution takes equal values at the extremes 0 and 1 (see Chianucci et al., 2018). For the uniform distribution, the parameters satisfy $\alpha=\beta=1$. Therefore, for plants exhibiting plagiophile, extremophile, or uniform leaf distributions, the transformed angle \( \theta = 2\theta_{leaf} \) can be treated as an axial angle, given the \(\pi\)-periodicity of its density function, and the NNTS axial models can be applied to model the distribution of \( \theta \).

\begin{figure}[h]
\center{
\includegraphics[scale=.9, bb= 0 0 504 504]{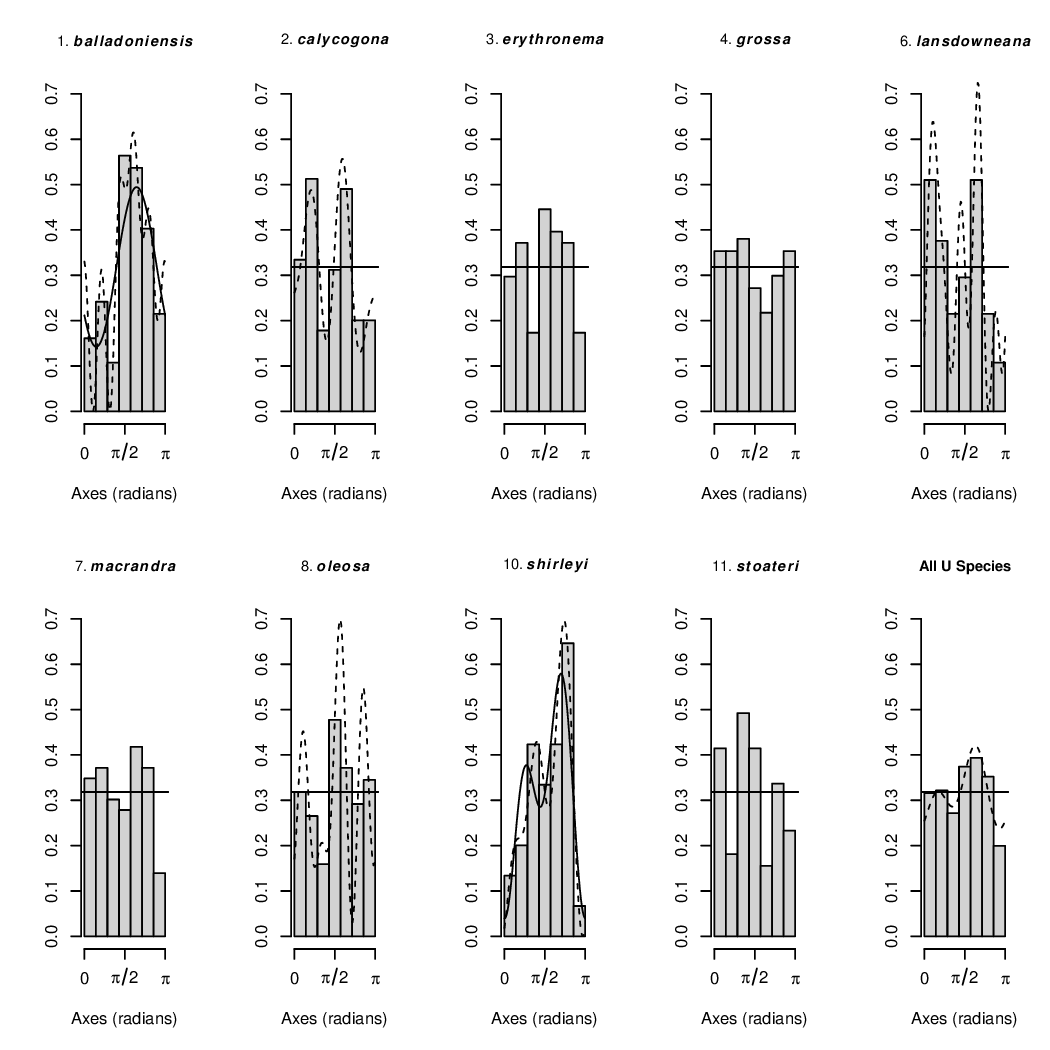}
\caption{Leaf inclination angles (uniform species - U): Histogram with best BIC-fitted (solid line) and best AIC-fitted (dashed line) NNTS axial densities.
\label{leafangleugraph}
}
}
\end{figure}

\begin{figure}[h]
\center{
\includegraphics[scale=0.7, bb= 0 0 504 504]{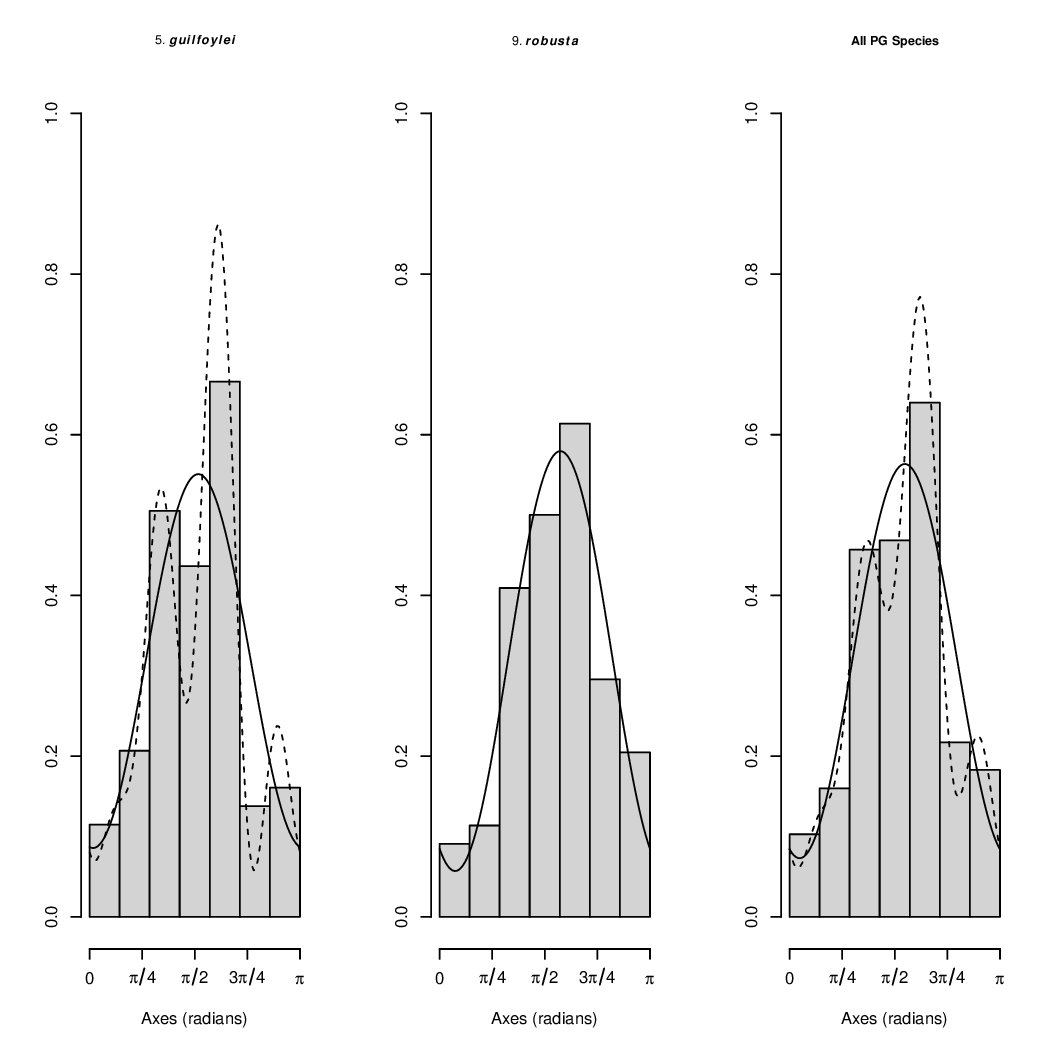}
\caption{Leaf inclination angles (plagiophile species - PG): Histogram with best BIC-fitted (solid line) and best AIC-fitted (dashed line) NNTS axial densities.
\label{leafanglepggraph}
}
}
\end{figure}

The dataset from Pisek \& Adamson (2020) contains leaf inclination angle measurements for 71 species of gum trees (genus \emph{Eucalyptus}). From these, we selected 11 species observed at the Huntington Library, Art Collections, and Botanical Gardens in Pasadena, California (latitude 34.125, longitude -118.114, altitude 207 m.a.s.l.), classified by Pisek \& Adamson (2020) as either uniform (9 species) or plagiophile (2 species).
Table \ref{leafstable} provides species names, sample sizes, Beta distribution parameter estimates, the optimal \(M\) values for the best-fitting NNTS axial models based on the Bayesian Information Criterion (BIC), and the corresponding BIC values. Figures \ref{leafangleugraph} and \ref{leafanglepggraph} display circular dot plots, histograms, and the best BIC-fitted NNTS axial densities for species classified as uniform (U) and plagiophile (PG), respectively.

Notably, for species classified as uniform (cases 1 \emph{balladoniensis} and 10 \emph{shirleyi}), neither the best BIC- nor AIC-selected NNTS axial densities corresponded to a uniform distribution (Figure \ref{leafangleugraph}). This observation is further confirmed by the uniformity test of Fern\'andez-Dur\'an \& Gregorio-Dom\'inguez (2024) shown in Table \ref{leafstable}, where the null hypothesis of uniformity is rejected for cases 1 and 10 when comparing against either a NNTS axial model with \(M=1\) or the best BIC-selected NNTS axial model. Moreover, when using the best AIC-selected NNTS axial model as an alternative, the hypothesis of uniformity is additionally rejected for cases 2, 6, and 8 at a 5\% significance level.

Table~\ref{leafstable2} presents the \( p \)-values of the homogeneity tests for the nine cases of U species (column~5), the two cases of PG species (column~9), and the combined eleven cases of U and PG species (column~13), for values of the parameter \( M_{\text{pooled}} \) ranging from~1 to~8. 
For the U species, the null hypothesis of homogeneity is clearly rejected, with Bonferroni- and Tippett-corrected \( p \)-values both equal to 2.6028e-07. For the PG species, the individual tests for \( M_{\text{pooled}} = 1, \ldots, 8 \) do not reject the null hypothesis of homogeneity. The Bonferroni- and Tippett-corrected \( p \)-values are equal to~1.0000 and~0.7924, respectively, indicating no rejection of the null hypothesis of homogeneity. 
Finally, for the combined U and PG species, as expected, both the individual and corrected \( p \)-values are very small, clearly rejecting the null hypothesis of homogeneity.

\section{Conclusions}

A flexible new family of axial data densities is developed by identifying conditions under which NNTS densities, originally defined for circular data, are suitable for axial data. These NNTS axial densities may be uniform, symmetric, asymmetric, or multimodal, depending on parameter constraints. Efficient numerical optimization on manifolds, adapted from Fern\'andez-Dur\'an \& Gregorio-Dom\'inguez (2010), facilitates maximum likelihood estimation and enables direct implementation of likelihood ratio tests. The proposed methodology was successfully applied to real datasets involving the orientations of rocks, animals, and plants, showing excellent performance. Importantly, these results demonstrate significant improvements over previous methodologies that did not consider multimodal axial densities.

\section*{Acknowledgments}

The authors wish to thank the Asociaci\'on Mexicana de Cultura, A.C. for its support and Prof. Sabine Begall for providing the dataset on magnetic orientations of ruminants.

\newpage

\renewcommand{\baselinestretch}{1.00}

\begin{table}[t]
\begin{center}
\scalebox{0.85}{
\begin{tabular}{c|ccc|ccc|ccc}
\hline
    & \multicolumn{3}{c|}{Feldspar laths in basalt}          & \multicolumn{3}{c|}{Feldspar laths in basalt} & \multicolumn{3}{c}{Face cleat}      \\
    & \multicolumn{3}{c|}{133 observations}                  & \multicolumn{3}{c|}{60 observations}          & \multicolumn{3}{c}{63 observations} \\
$M$ & loglik    & BIC     & AIC                              & loglik    & BIC     & AIC & loglik    & BIC     & AIC \\
\hline
0   & -152.25  & 304.50*  & 304.50  &  -68.68 & 137.37* & 137.37  & -72.12  & 144.24 & 144.24  \\
1   & -150.36  & 310.49   & 304.71  &  -65.12 & 138.42  & 134.23* & -34.78  &  77.85 &  73.57  \\
2   & -147.66  & 314.89   & 303.33  &  -64.62 & 145.61  & 137.24  & -19.80  &  56.17 &  47.60  \\
3   & -143.33  & 316.01   & 298.66  &  -64.58 & 153.72  & 141.16  & -15.31  &  55.47 &  42.61  \\
4   & -137.26  & 313.64   & 290.51* &  -63.59 & 159.93  & 143.17  &  -8.71  &  50.56 &  33.42  \\
5   & -136.90  & 322.70   & 293.80  &  -62.95 & 166.84  & 145.90  &   3.49  &  34.45 &  13.01  \\
6   & -136.61  & 331.90   & 297.21  &  -61.54 & 172.22  & 147.08  &  10.10  &  29.52*&   3.80* \\
7   & -134.28  & 337.03   & 296.56  &         &         &         &  11.95  &  34.11 &   4.10  \\
8   & -132.50  & 343.25   & 297.00  &         &         &         &         &        &         \\
9   & -132.13  & 352.29   & 300.27  &         &         &         &         &        &         \\
10  & -130.35  & 358.50   & 300.70  &         &         &         &         &        &         \\
11  & -129.24  & 366.08   & 302.49  &         &         &         &         &        &         \\
12  & -128.94  & 375.25   & 305.88  &         &         &         &         &        &         \\
13  & -128.41  & 383.97   & 308.82  &         &         &         &         &        &         \\
\hline
\end{tabular}}
\renewcommand{\baselinestretch}{1}
\caption{Columns 2 to 4: Feldspar laths in basalt dataset with 133 observations: Maximized log-likelihood (loglik), BIC, and AIC values for NNTS axial models with \(M=0\) to 13. Columns 5 to 7: Feldspar laths in basalt dataset with 60 observations: \(M\), maximized log-likelihood (loglik), BIC, and AIC values for NNTS axial models with \(M=0\) to 6. Columns 8 to 10: Face cleat dataset with 63 observations: \(M\), Maximized log-likelihood (loglik), BIC, and AIC values for NNTS axial models with \(M=0\) to 7. Best BIC and AIC models are marked with an asterisk.
\label{feldspar0tablethreetogether} }
\end{center}
\end{table}

\begin{table}[t]
\begin{center}
\scalebox{0.63}{
\begin{tabular}{c|ccc|c}
\hline
    & \multicolumn{3}{c|}{Feldspar laths in basalt}          & homogeneity \\
$M_{pooled}$  & loglik    & BIC     & AIC                    & $p$-value     \\
\hline
0   & -220.93  & 441.87* & 441.87  &  \\
1   & -216.27  & 443.06  & 436.54  & 0.4514 \\
2   & -214.56  & 450.18  & 437.13  & 0.3353 \\
3   & -211.27  & 454.11  & 434.54  & 0.3480 \\
4   & -206.63  & 455.37  & 429.27* & 0.1709 \\
5   & -205.92  & 464.46  & 431.83  & 0.2759 \\
6   & -205.88  & 474.91  & 435.76  & 0.2171 \\
\hline
\end{tabular}}
\renewcommand{\baselinestretch}{1}
\caption{Feldspar laths in basalt pooled dataset with 193 observations (133 from dataset~B2 and 60 from dataset~B5, both from Fisher,~1993): 
\(M_{pooled}\), maximized log-likelihood (loglik), BIC, and AIC values for the fitted NNTS axial models. 
The best models according to the BIC and AIC criteria are indicated with an asterisk. 
The last column reports the \( p \)-values from the NNTS axial homogeneity test.
\label{feldsparpooled} }
\end{center}
\end{table}

\begin{table}[t]
\begin{center}
\scalebox{0.85}{
\begin{tabular}{cccccccccc}
\hline
Site & $n$    & Latitude     & Longitude     &  $M_{BIC}$    &  BIC     & $M_{AIC}$ & AIC & loglik $M_{BIC}$ & loglik $M =$ 5   \\
\hline
1   & 100 & -15$^\circ$ 43$''$ & 144$^\circ$ 42$''$ & 2   & 122.12   & 2   & 111.69 &  -51.85  &  -50.10    \\
2   &  50 & -15$^\circ$ 32$''$ & 144$^\circ$ 17$''$ & 3   &  42.48   & 4   &  30.50 &   -9.50  &   -6.92    \\
3   &  50 & -14$^\circ$ 59$''$ & 143$^\circ$ 35$''$ & 3   &  30.02   & 4   &  17.40 &   -3.27  &   -0.17    \\
4   &  50 & -14$^\circ$ 19$''$ & 143$^\circ$ 19$''$ & 4   &  33.12   & 4   &  17.82 &   -0.91  &    0.15    \\
5   &  50 & -13$^\circ$ 21$''$ & 142$^\circ$ 53$''$ & 3   &  37.56   & 3   &  26.09 &   -7.04  &   -3.69    \\
6   &  50 & -12$^\circ$ 50$''$ & 142$^\circ$ 44$''$ & 4   &  17.96   & 6   &   2.52 &    6.67  &    8.74    \\
7   &  66 & -11$^\circ$ 54$''$ & 142$^\circ$ 30$''$ & 4   &  46.92   & 4   &  29.40 &   -6.70  &   -5.69    \\
8   &  48 & -12$^\circ$ 06$''$ & 142$^\circ$ 33$''$ & 3   &  38.75   & 3   &  27.52 &   -7.76  &   -4.63    \\
9   & 100 & -12$^\circ$ 29$''$ & 142$^\circ$ 39$''$ & 3   &  66.85   & 3   &  51.22 &  -19.61  &  -17.65    \\
10  &  50 & -13$^\circ$ 12$''$ & 142$^\circ$ 46$''$ & 4   &   6.74   & 7   & -13.26 &   12.28  &   16.13    \\
11  &  37 & -15$^\circ$ 02$''$ & 143$^\circ$ 41$''$ & 4   &  13.37   & 6   &  -2.14 &    7.76  &   10.38    \\
12  &  31 & -14$^\circ$ 47$''$ & 143$^\circ$ 30$''$ & 4   &  28.27   & 5   &  15.49 &   -0.40  &    2.26    \\
13  & 132 & -13$^\circ$ 50$''$ & 143$^\circ$ 12$''$ & 5   &  -4.02   & 6   & -33.02 &   26.42  &   26.42    \\
14  &  92 & -13$^\circ$ 50$''$ & 143$^\circ$ 12$''$ & 4   &  25.36   & 5   &   2.98 &    5.41  &    8.51    \\
\hline
all & 906 &                    &                    & 5   & 274.43   & 6   & 225.31 & -103.17  & -103.17    \\
\hline
\end{tabular}}
\renewcommand{\baselinestretch}{1}
\caption{Termite mounds dataset: Site ID, sample size \(n\), and geographic coordinates (latitude, longitude) for each of the 14 sites are shown in the first four columns. Column 5 lists the optimal \(M\) according to BIC, and Column 6 the corresponding best BIC values. Column 7 lists the optimal \(M\) according to AIC, and Column 8 the corresponding best AIC values. Columns 9 and 10 report the maximized log-likelihood values for NNTS axial models with \(M=M_{BIC}\) and \(M=5\), respectively.
\label{termitestable} }
\end{center}
\end{table}

\begin{table}[t]
\begin{center}
\scalebox{0.63}{
\begin{tabular}{c|ccc|c||ccc|c||ccc|c}
\hline
              & \multicolumn{3}{c|}{Termite mounds (14 sites)}         & homogeneity & \multicolumn{3}{c|}{Sites 5 and 14}    & homogeneity & \multicolumn{3}{c|}{Sites 6 and 8}    & homogeneity \\
$M_{pooled}$  & loglik    & BIC     & AIC                              & $p$-value   & loglik    & BIC     & AIC              & $p$-value & loglik    & BIC     & AIC             & $p$-value   \\
\hline
0   & -1037.13 & 2074.25 & 2074.25 &         & -162.55  & 325.10 &  325.10 &          & -112.18 & 224.37 &  224.37 &          \\
1   &  -492.04 &  997.69 &  988.07 & 0.9951  &  -73.13  & 156.17 &  150.26 & 0.6330   &  -53.05 & 115.28 &  110.11 & 0.9913   \\
2   &  -255.54 &  538.32 &  519.08 & 0.8124  &  -31.19  &  82.21 &   70.39 & 0.6345   &  -25.22 &  68.77 &   58.43 & 0.7750   \\
3   &  -151.17 &  343.19 &  314.33 & 0.1904  &  -12.97  &  55.67 &   37.93 & 0.5923   &   -6.53 &  40.58 &   25.07 & 0.9873   \\
4   &  -116.25 &  286.98 &  248.51 & 0.0050  &   -4.09  &  47.83*&   24.19 & 0.4428   &   -1.63 &  39.93*&   19.25 & 0.8652   \\
5   &  -103.17 &  274.43*&  226.33 & 0.0062  &    0.38  &  48.80 &   19.24*& 0.5431   &    1.12 &  43.62 &   17.77*& 0.8160   \\
6   &  -100.66 &  283.02 &  225.31*& 0.0204  &    1.92  &  55.63 &   20.16 & 0.6819   &    1.60 &  51.83 &   20.81 & 0.7895   \\
7   &  -100.19 &  295.71 &  228.38 & 0.0448  &    2.12  &  65.14 &   23.76 & 0.8098   &    1.91 &  60.38 &   24.19 & 0.8791   \\
8   &   -99.68 &  308.31 &  231.36 & 0.1424  &    2.56  &  74.17 &   26.87 & 0.8735   &    2.91 &  67.54 &   26.18 & 0.9058   \\
\hline
\end{tabular}}
\renewcommand{\baselinestretch}{1}
\caption{Orientations of termite mounds pooled dataset with 906 observations (14 datasets measured at different latitudes and longitudes; see Table~\ref{termitestable}), pooled dataset of sites~5 and~14 (142 observations), and pooled dataset of sites~6 and~8 (98 observations): \(M_{pooled}\),
maximized log-likelihood (loglik), BIC, and AIC values for the fitted NNTS axial models. 
The best models according to the BIC and AIC criteria are indicated with an asterisk. 
Columns 5, 9, and 13 report the \(p\)-values from the corresponding NNTS axial homogeneity tests.
\label{termitespooled} }
\end{center}
\end{table}

\begin{table}[t]
\begin{center}
\scalebox{0.63}{
\begin{tabular}{c|ccc|ccc|ccc}
\hline
    & \multicolumn{3}{c|}{Cattle}          & \multicolumn{3}{c|}{Red Deer} & \multicolumn{3}{c}{Roe Deer} \\
$M$ & loglik    & BIC     & AIC            & loglik    & BIC     & AIC    & loglik    & BIC     & AIC      \\
\hline
0   & -352.58  & 705.15  & 705.15  & -45.79  & 91.58  & 91.58  & -230.09  & 460.18  & 460.18  \\
1   & -275.77  & 562.99* & 555.53  & -21.46  & 50.31  & 46.93  & -123.40  & 257.40  & 250.79  \\
2   & -272.92  & 568.76  & 553.84* & -10.40  & 35.56  & 28.80  &  -70.66  & 162.53  & 149.31  \\
3   & -271.25  & 576.88  & 554.50  &  -2.41  & 26.96* & 16.82  &  -42.29  & 116.40  &  96.58  \\
4   & -270.30  & 586.43  & 556.59  &   0.88  & 27.74  & 14.23  &  -32.23  & 106.88  &  80.45  \\
5   & -268.71  & 594.73  & 557.43  &   4.25  & 28.38  & 11.49  &  -26.76  & 106.56  &  73.53  \\
6   & -268.09  & 604.94  & 560.18  &   6.58  & 31.11  & 10.85* &  -18.43  & 100.50* &  60.86* \\
7   & -267.85  & 615.92  & 563.70  &         &        &        &  -17.87  & 109.99  &  63.75  \\
8   & -267.77  & 627.22  & 567.54  &         &        &        &  -17.50  & 119.86  &  67.00  \\
9   & -266.00  & 635.15  & 568.01  &         &        &        &  -15.44  & 126.34  &  66.88  \\
10  & -265.66  & 645.91  & 571.31  &         &        &        &  -14.87  & 135.81  &  69.75  \\
\hline
\end{tabular}}
\renewcommand{\baselinestretch}{1}
\caption{Ruminant magnetic orientations (308 observations for cattle, 40 for red deer, and 201 for roe deer): \(M\), Maximized log-likelihood (loglik), BIC, and AIC values for fitted NNTS axial models. Best BIC and AIC models are marked with an asterisk.
\label{rumianttable} }
\end{center}
\end{table}

\begin{table}[t]
\begin{center}
\scalebox{0.63}{
\begin{tabular}{c|ccc|c|ccc|c}
\hline
             &  \multicolumn{3}{c|}{Cattle, Red, Roe Deer} & homogeneity & \multicolumn{3}{c|}{Red, Roe Deer} & homogeneity \\
$M_{pooled}$ &  loglik    & BIC     & AIC  & $p$-value  & loglik    & BIC     & AIC & $p$-value   \\
\hline
0   & -628.46  & 1256.91  & 1256.91 &             & -275.88 & 551.76 & 551.76 &   \\
1   & -424.91  &  862.43  &  853.81 & 0.0731      & -145.37 & 301.71 & 294.74 & 0.6002  \\
2   & -380.34  &  785.92  &  768.69 & 1.2125e-08  &  -81.44 & 184.82 & 170.88 & 0.9432  \\
3   & -370.96  &  779.77  &  753.92 & 5.9395e-18  &  -45.68 & 124.28 & 103.37 & 0.9231  \\
4   & -361.87  &  774.20* &  739.74 & 4.4756e-18  &  -34.13 & 112.14 &  84.27 & 0.6939  \\
5   & -356.94  &  776.95  &  733.87*& 2.1024e-18  &  -25.57 & 105.99 &  71.14 & 0.8047  \\
6   & -355.46  &  786.62  &  734.92 & 2.1300e-20  &  -14.58 &  94.97*&  53.15*& 0.9415  \\
7   & -354.64  &  797.59  &  737.28 & 5.2502e-19  &  -13.60 & 103.99 &  55.21 & 0.9738  \\
8   & -354.02  &  808.96  &  740.03 & 8.9523e-18  &  -13.00 & 113.75 &  58.00 & 0.9765  \\
9   & -352.15  &  817.85  &  740.31 & 2.5878e-17  &  -10.70 & 120.13 &  57.40 & 0.9837  \\
10  & -351.43  &  829.02  &  742.86 & 3.4098e-16  &   -9.94 & 129.58 &  59.88 & 0.9931  \\
\hline
\end{tabular}}
\renewcommand{\baselinestretch}{1}
\caption{Ruminant magnetic orientations (549 observations for cattle, red deer, and roe deer, and 241 observations for red and roe deer): \(M_{pooled}\), maximized log-likelihood (loglik), BIC, and AIC values for fitted NNTS axial models. Best BIC and AIC models are marked with an asterisk.
Columns 5 and 9 report the \(p\)-values from the corresponding NNTS axial homogeneity tests.
\label{rumianttable2} }
\end{center}
\end{table}

\begin{table}[t]
\begin{center}
\scalebox{0.85}{
\begin{tabular}{llccccccccccc}
\hline
id & Species name          & $n$    & \multicolumn{2}{c}{Beta dist.} & Type & $M_{BIC}$ &  $M_{AIC}$  & \multicolumn{3}{c}{$p$-value unif. test vs.} & \multicolumn{2}{c}{loglik}   \\
   & \emph{Eucalyptus ...} &        & $\hat{\alpha}$     & $\hat{\beta}$    &      &           &             & $M_{BIC}$  & $M_{AIC}$  & $M$=1            & $M_{BIC}$ & $M_{AIC}$ \\
\hline
1  & \emph{balladoniensis} &  83 & 1.62 & 1.31 & U  & 1  & 5  & <0.01 & <0.01        & <0.01  & -87.59 &  -77.71      \\
2  & \emph{calycogona}     & 100 & 1.06 & 1.29 & U  & 0  & 3  &       & <0.01        & >0.10  &-114.47 & -105.64      \\
3  & \emph{erythronema}    &  90 & 1.18 & 1.23 & U  & 0  & 0  &       &              & >0.10  &-103.03 & -103.03      \\
4  & \emph{grossa}         &  82 & 0.81 & 0.85 & U  & 0  & 0  &       &              & >0.10  & -93.87 &  -93.87      \\
6  & \emph{lansdowneana}   &  83 & 0.96 & 1.29 & U  & 0  & 5  &       & <0.01        & >0.10  & -95.01 &  -79.00      \\
7  & \emph{macrandra}      &  96 & 1.04 & 1.20 & U  & 0  & 0  &       &              & >0.10  &-109.89 & -109.89      \\
8  & \emph{oleosa}         &  84 & 1.07 & 1.01 & U  & 0  & 4  &       & >0.01        & >0.10  & -96.16 &  -86.58      \\
   &                       &     &      &      &    &    &    &       & <0.05        &        &        &              \\
10 & \emph{shirleyi}       & 100 & 1.90 & 1.61 & U  & 2  & 3  & <0.01 & <0.01        & <0.01  &-100.39 &  -96.85      \\
11 & \emph{stoateri}       &  86 & 1.02 & 1.15 & U  & 0  & 0  &       &              & >0.10  & -98.45 &  -98.45      \\
\hline
   & \emph{All U Species}  & 804 &      &      &    & 0  & 2  &       & <0.01        & <0.01  &-920.36 & -909.67      \\
\hline
5  & \emph{guilfoylei}     &  97 & 2.29 & 2.19 & PG & 1  & 4  & <0.01 & <0.01        & <0.01  & -95.95 &  -87.69      \\
9  & \emph{robusta}        &  98 & 2.64 & 2.13 & PG & 1  & 1  & <0.01 & <0.01        & <0.01  & -93.24 &  -93.24      \\
\hline
   & \emph{All PG Species} & 195 &      &      &    & 1  & 4  & <0.01 & <0.01        & <0.01  &-190.51 & -183.90      \\
\hline
\end{tabular}}
\renewcommand{\baselinestretch}{1}
\caption{Leaf inclination angles dataset: For each \emph{Eucalyptus} species, the table provides sample size \(n\), fitted Beta distribution parameter estimates \(\hat{\alpha}\) and \(\hat{\beta}\), theoretical leaf orientation type (U for uniform, PG for plagiophile), optimal \(M\) for the best-fitting NNTS axial model based on BIC, corresponding BIC values, maximized log-likelihood values (loglik), the uniformity test $p$-value from Fern\'andez-Dur\'an \& Gregorio-Dom\'inguez (2024), and the maximized log-likelihood values for the NNTS axial models with $M_{BIC}$ and $M_{AIC}$.  
\label{leafstable} }
\end{center}
\end{table}

\begin{table}[t]
\begin{center}
\scalebox{0.63}{
\begin{tabular}{c|ccc|c|ccc|c|ccc|c}
\hline
             &  \multicolumn{3}{c|}{All U species}    & homogeneity & \multicolumn{3}{c|}{All PG species} & homogeneity & \multicolumn{3}{c|}{All U and All PG species} & homogeneity \\
$M_{pooled}$ &  loglik    & BIC     & AIC  & $p$-value& loglik    & BIC     & AIC & $p$-value & loglik    & BIC     & AIC & $p$-value  \\
\hline
0   & -920.36  & 1840.73  & 1840.73   &              & -223.22  & 446.44  & 446.44  &         & -1143.59   & 2287.17   & 2287.17   &            \\
1   & -914.51  & 1842.40* & 1833.02   & 0.0091       & -190.51  & 391.56* & 385.02* & 0.2697  & -1122.36   & 2258.53*  & 2248.72   & 1.1723e-08 \\
2   & -909.67  & 1846.10  & 1827.34*  & 4.9449e-06   & -190.27  & 401.63  & 388.54  & 0.5062  & -1116.94   & 2261.51   & 2241.89*  & 1.0143e-11 \\
3   & -908.81  & 1857.76  & 1829.62   & 1.8418e-06   & -187.52  & 406.67  & 387.03  & 0.2820  & -1116.94   & 2275.32   & 2245.88   & 2.1859e-13 \\
4   & -907.96  & 1869.44  & 1831.93   & 6.4032e-09   & -183.90  & 409.99  & 383.81  & 0.3602  & -1116.54   & 2288.33   & 2249.08   & 3.2160e-17 \\
5   & -905.45  & 1877.81  & 1830.91   & 3.2535e-08   & -183.60  & 419.94  & 387.21  & 0.2843  & -1114.54   & 2298.15   & 2249.08   & 1.3807e-16 \\
6   & -905.44  & 1891.15  & 1834.88   & 7.5061e-08   & -182.76  & 428.80  & 389.53  & 0.2534  & -1114.44   & 2311.77   & 2252.89   & 2.7303e-16 \\
7   & -902.89  & 1899.43  & 1833.78   & 1.7020e-07   & -182.44  & 438.70  & 392.88  & 0.2209  & -1112.69   & 2322.07   & 2253.38   & 4.0274e-16 \\
8   & -902.36  & 1911.76  & 1836.73   & 8.7638e-07   & -181.83  & 448.02  & 395.65  & 0.1784  & -1112.23   & 2334.97   & 2256.46   & 2.0702e-15 \\
\hline
\end{tabular}}
\renewcommand{\baselinestretch}{1}
\caption{Leaf inclination angles pooled datasets (all U species with 804 observations, all PG species with 195 observations, and all U and PG species combined with 999 observations). Values of \( M_{\text{pooled}} \) for the NNTS axial model, maximized log-likelihood (loglik), and the corresponding BIC and AIC values.
Columns 5, 9, and 13 report the \(p\)-values from the corresponding NNTS axial homogeneity tests.
\label{leafstable2} }
\end{center}
\end{table}

\newpage

\thebibliography{99}

\bibitem{1} Abe, T. Shimizu, K., Kuuluvainen, T. \& Aakala, T. (2012) Sine-skewed axial distributions with an application to fallen tree data. \emph{Environmental and Ecological Statistics}, 19, 295-307.
        
\bibitem{2} Arnold, B.C. \& SenGupta, A. (2006) Probability distributions and statistical inference for axial data. \emph{Environmental and Ecological Statistics}, 13, 271-285.

\bibitem{3} Arnold, B.C. \& SenGupta, A. (2011) Models for axial data. In: Wells, M., SenGupta, A. (Eds.) \emph{Advances in Directional and Linear Statistics}, Physica-Verlag HD. pp. 1-9. 

\bibitem{4} Barlow D.J. \& Thornton J.M. (1988) Helix geometry in proteins. \emph{Journal of Molecular Biology}, 201, 601-619. 

\bibitem{5} Batschelet, E. (1981) \emph{Circular Statistics in Biology}. New York:Academic Press.

\bibitem{6} Begall, S., \v{C}erven\'y, J., Neef, J., Vojt\v{e}ch, O. \& Burda, H. (2008) Magnetic alignment in grazing and resting cattle and deer. \emph{Proceedings of the National Academy of Sciences}, 105, 13451-13455.

\bibitem{7} Chianucci, F., Pisek, J., Raabe, K., Marchino, L., Ferrara, C. \& Corona, P. (2018) A dataset of leaf inclination angles for temperate and boreal broadleaf woody species.
\emph{Annals of Forest Science}, 75:50.

\bibitem{8} Cinar, O. \& Viechtbauer W. (2022) The poolr package for combining independent and dependent p-values. \emph{Journal of Statistical Software}, 101, 1-42.

\bibitem{9} de Wit, C.T. (1965) Photosyntesis of leaf canopies. \emph{Agricultural Research Report}, no. 663, Wageningen.

\bibitem{10} Fern\'andez-Dur\'an, J.J. \& Gregorio-Dom\'inguez, M.M. (2010) Maximum likelihood estimation of nonnegative trigonometric sums models using a Newton-like algorithm on manifolds. \emph{Electronic Journal of Statistics}, 4, 1402-10.

\bibitem{11} Fern\'andez-Dur\'an, J.J. \& Gregorio-Dom\'inguez, M.M. (2012) CircNNTSR: an R package for the statistical analysis of circular data using nonnegative trigonometric sums (NNTS) models. Available from: http://CRAN.R-project.org/package=CircNNTSR. R package version 2.0. 

\bibitem{12} Fern\'andez-Dur\'an, J. J. \& Gregorio-Dom\'inguez, M. M. (2024) Sums of independent circular random variables and maximum likelihood circular uniformity tests based on 
nonnegative trigonometric sums distributions. \emph{AppliedMath}, 4, 495-516. Available from: https://doi.org/10.3390/appliedmath4020026.

\bibitem{13} Fisher, N.I. (1993) \emph{Statistical Analysis of Circular Data}. Cambridge, New York: Cambridge University Press.

\bibitem{14} Fitak, R.R. \& S\"onke, J. (2017) Bringing the analysis of animal orientation data full circle: model-based approaches with maximum likelihood. \emph{Journal of Experimental Biology}, 220, 3878-3882.

\bibitem{15} Goeman, J.J. \& Solari, A. (2014). Multiple hypothesis testing in genomics. \emph{Statistics in Medicine}, 33, 1946-1978.

\bibitem{16} Hawley, D.L. \& Peebles, P.J.E. (1975) Distribution of observed orientations of galaxies. \emph{The Astronomical Journal}, 80, 477-491.

\bibitem{17} Koch, D.D., Ali, S.F., Weikert, M.P., Shirayama, M., Jenkins, R. \& Wang, L. (2012) Contribution of posterior corneal astigmatism to total corneal astigmatism. \emph{Journal of Cataract \& Refractive Surgery}, 38, 2080-2087.

\bibitem{18} Mingione, M., Lagona, F., Nagar, P., von Holtzhausen, F., Bekker, A., Schoombie, J. \& le Roux, P. C. (2025) Does wind affect the orientation of vegetation stripes? A copula-based mixture model for axial and circular data. \emph{Environmetrics}, 36, e70021. 
  
\bibitem{19} Mardia, K.V. \& Jupp, P.E. (2000) \emph{Directional Statistics}. Chichester, New York: John Wiley and Sons.

\bibitem{20} Onstott, T. (1980) Application of the Bingham distribution function in paleomagnetic studies. \emph{Journal of Geophysical Research: Solid Earth}, 85, 1500-1510. 

\bibitem{21} Pisek, J. \& Adamson, K. (2020) Dataset of leaf inclination angles for 71 different \emph{Eucalyptus} species. \emph{Data in Brief}, 33, 106391

\bibitem{22} R Development Core Team (2012) R: a language and environment for statistical computing. R Foundation for Statistical Computing, Vienna, Austria. ISBN 3-900051-07-0. Available from: http://www.R-project.org/.

\bibitem{23} Sherratt, J. A. (2015) Using wavelength and slope relationships to infer the historical origin of semiarid vegetation bands. \emph{Proceedings of the National Academy of Sciences of the United States of America}, 112, 4202-4207. 
  
\bibitem{24} Tauxe, L. (2010) \emph{Essentials of Paleomagnetism}. Berkeley, Los Angeles, London: University of California Press.

\bibitem{25} Vrtovec, T., Pernu\v{s}, F. \& Likar, B. (2009) A review of methods for quantitative evaluation of axial vertebral rotation. \emph{European Spine Journal}, 18, 1079-1090.

\bibitem{26} Wiltschko, R. \& Wiltschko, W. (1995) Magnetic orientation in animals. Berlin, Heidelberg:Springer-Verlag.

\bibitem{27} Wiltschko, W. \& Wiltschko, R. (2005) Magnetic orientation and magnetoreception in birds and other animals. \emph{Journal of Comparative Physiology A}, 191, 675-693.
\end{document}